\documentclass[review]{elsarticle}
\usepackage{graphicx}
\usepackage{amsmath}
\usepackage{amssymb}
\usepackage{bm}
\usepackage{subfigure}
\usepackage{color}

\title{GPU-based Swendsen-Wang multi-cluster algorithm for the simulation 
of two-dimensional classical spin systems} 
\author[tmu]{Yukihiro Komura}
\ead{y-komura@phys.se.tmu.ac.jp}
\author[tmu]{Yutaka Okabe}
\ead{okabe@phys.se.tmu.ac.jp}
\address[tmu]{Department of Physics, Tokyo Metropolitan University, Hachioji, Tokyo 192-0397, Japan}

\begin{document}

\begin{abstract}
We present the GPU calculation with the common unified device architecture 
(CUDA) for the Swendsen-Wang multi-cluster algorithm 
of two-dimensional classical spin systems. 
We adjust the two connected component labeling algorithms 
recently proposed with CUDA for the assignment of the cluster 
in the Swendsen-Wang algorithm.  
Starting with the $q$-state Potts model, we extend our implementation 
to the system of vector spins, the $q$-state clock model, 
with the idea of embedded cluster.  
We test the performance, and the calculation time on GTX580 
is obtained as 2.51 nano sec per a spin flip for the $q=2$ Potts model 
(Ising model) and 2.42 nano sec per a spin flip for the $q=6$ clock model 
with the linear size $L=4096$ at the critical temperature, respectively. 
The computational speed for the $q=2$ Potts model 
on GTX580 is 12.4 times as fast as the calculation speed 
on a current CPU core. 
That for the $q=6$ clock model 
on GTX580 is 35.6 times as fast as the calculation speed 
on a current CPU core. 
\end{abstract}

\begin{keyword}
 Monte Carlo simulation \sep
 cluster algorithm \sep
 Ising model \sep
 parallel computing \sep
 GPU 
\end{keyword}

\maketitle

\section{Introduction}
Computer simulation is an essential tool for studying physical properties 
of many-particle systems. The Metropolis-type Monte Carlo simulation 
\cite{metro53} with a single spin flip has been a success as a standard method 
of simulation of many-particle systems.  However, the single-spin-flip 
algorithm often suffers from the problem of slow dynamics 
or the critical slowing down; that is, the relaxation time 
diverges at the critical temperature. 
To overcome difficulty, a cluster flip algorithm was proposed 
by Swendsen and Wang \cite{sw87}.  
They applied the Fortuin-Kasteleyn \cite{fk72} 
representation to identify clusters of spins.  
The problem of the thermal phase transition is mapped onto 
the geometric percolation problem in the cluster formalism.  
In the cluster algorithm, spins in the cluster are updated 
at a time.  In the Swendsen-Wang (SW) algorithm, all the spins 
are partitioned into clusters; thus, the SW algorithm is 
called the multi-cluster algorithm. 
Wolff \cite{wolff89} proposed another type of cluster algorithm, 
that is, a single-cluster algorithm, where only a single cluster 
is generated, and the spins of that cluster are updated. 
Although the cluster algorithm was originally formulated 
for the scalar order parameter, such as the Potts model, 
Wolff \cite{wolff89} introduced the idea of embedded cluster 
to deal with systems of vector spins, such as 
the classical XY model or the classical Heisenberg model. 

Computational physics develops with the advance in computer technology. 
Recently the use of general purpose computing on graphics processing 
unit (GPU) is a hot topic in computer science. 
Drastic reduction of processing times can be realized in 
scientific computations. 
Using the common unified device architecture (CUDA) released by NVIDIA, 
it is now easy to implement algorithms on GPU 
using standard C or C++ language with CUDA specific extension.  

Preis {\it et al.} \cite{preis09} studied the two-dimensional (2D) and 
three-dimensional (3D) Ising models 
by using the Metropolis algorithm with CUDA.  
They used a variant of sublattice decomposition 
for a parallel computation on GPU. 
The spins on one sublattice do not interact with other spins 
on the same sublattice.  Therefore one can update all spins 
on a sublattice in parallel when making the Metropolis simulation. 
As a result they were able to accelerate 60 times for the 2D Ising model 
and 35 times for the 3D Ising model compared to a current CPU core. 
Recently, the GPU acceleration of the multispin coding 
of the Ising model was discussed \cite{block10}. 
Moreover, many attempts for simulating lattice spin models on GPU 
using the Metropolis algorithm were reported 
\cite{Levy,Bernaschi_GPU_spin_glass,weigel_spin_model}. 

Since the Metropolis algorithm has the problem of slow dynamics 
as mentioned above, and this problem becomes conspicuous 
with increasing the system size, it is highly desirable 
to apply the GPU-based calculation to cluster algorithms. 
Only limited trials have been reported so far.  
The present authors \cite{komura11} have proposed the GPU-based 
calculation with CUDA for the Wolff single-cluster algorithm, 
where parallel computations are performed for the newly added spins 
in the growing cluster. 
Hawick {\it et al.} \cite{Hawick_single_cluster} have studied 
the CUDA implementation of the Wolff algorithm 
using a modified connected component labeling 
for the assignment of the cluster.  They put more emphasis on 
the hybrid implementation of Metropolis and Wolff updates and 
the optimal choice of the ratio of both updates. 
Quite recently, Weigel \cite{weigel11} has studied parallelization of 
cluster labeling and cluster update algorithms for calculations 
with CUDA. He realized the SW multi-cluster algorithm 
by using the combination of self-labeling algorithm and 
label relaxation algorithm or hierarchical sewing algorithm. 

In this paper, we present the GPU-based calculation with CUDA 
for the SW multi-cluster algorithm of 2D classical spin systems. 
We realize the SW cluster algorithm by using the connected component 
labeling algorithm for the assignment of clusters. 
The rest of the paper is organized as follows. 
In section 2, we briefly describe the standard way of implementing 
the SW algorithm on CPU. 
In section 3, we explain two types of connected component labeling 
which are used in the present calculation, and the idea of 
implementing the SW cluster algorithm on GPU. 
In section 4, we compare the performance of GPU calculation 
with that of CPU calculation. 
The summary and discussion are given in section 5. 

\section{Swendsen-Wang cluster algorithm}
We start with the Potts model whose Hamiltonian is given by
\begin{equation}
 \mathcal{H} = -J \sum_{<i,j>}(\delta_{S_{i},S_{j}}-1), 
              \quad S_{i} = 1, 2, \cdots, q, 
\end{equation}
and for $q$ = 2 this corresponds to the Ising model.
Here, $J$ is the coupling and $S_{i}$ is the Potts spin 
on the lattice site $i$. The summation is taken over 
the nearest neighbor pairs $<i,j>$.  
Periodic boundary conditions are employed. 

Swendsen and Wang proposed a Monte Carlo algorithm 
of multi-cluster flip \cite{sw87}. 
There are three main steps in the SW algorithm:
(1) Construct a bond lattice of active or non-active bonds. 
(2) The active bonds partition the spins into clusters which are 
identified and labeled using a cluster-labeling algorithm. 
(3) All spins in each cluster are set randomly to one of $q$.
The cluster identification problem is a variant of 
connected component labeling, which is an algorithmic 
application of graph theory.
For an efficient cluster-labeling algorithm, 
the Hoshen-Kopelman algorithm \cite{Hoshen_Kopelman}, which was 
first introduced in context of cluster percolation, is 
often used. The Hoshen-Kopelman algorithm 
is a special version of the class of union-and-find algorithms 
\cite{cormen}, and
has an advantage over 
other methods in low computer memory usage and 
short computational time. 

The actual spin-update process of the SW cluster algorithm 
on a CPU can be formulated as follows \cite{janke,landau}: 
\begin{itemize}
\item[(i)] Choose a site $i$. 
\item[(ii)] Look at each of the nearest neighbors $j$. If $S_j$ 
is equal to $S_i$, generate bond between site $i$ and $j$ 
with probability $p=1-e^{-\beta}$, where $\beta$ is 
the inverse temperature $J/T$. 
\item[(iii)] 
Choose the next spin and go to (i) until all sites are checked. 
\item[(iv)] 
Apply the Hoshen-Kopelman algorithm \cite{Hoshen_Kopelman} to identify all clusters. 
\item[(v)] 
Choose a cluster. 
\item[(vi)] 
Assign the spins $S_i$ in the cluster to one of $q$ 
with probability $1/q$. 
\item[(vii)] 
Choose another cluster and go to (vi) until all clusters are checked. 
\item[(viii)] Go to (i).
\end{itemize}
The procedures from (i) to (iii) correspond to the step of 
active bond generation.  
The procedure (iv) corresponds to the step of cluster labeling.
Those from (v) to (vii) correspond 
to the step of spin flip. 

In the Hoshen-Kopelman cluster-labeling algorithm, integer labels 
are assigned to each spin in a cluster. Each cluster has 
its own distinct set of labels.  The proper label of a cluster, 
which is defined to be the smallest label of any spin in the cluster, 
is found by the following function. 
The array \verb+label+ is used, and 
if \verb+label+ is a label belonging to a cluster, 
the \verb+label[label]+ is the index of another label 
in the same cluster which has a smaller value 
if such a smaller value exists.  The proper label for the cluster 
is found by evaluating \verb+label[label]+ repeatedly.

\section{GPU calculation of the Swendsen-Wang cluster algorithm}
Since the calculations of the step of active bond generation 
and the step of spin flip are done independently on each site, 
these steps are well suited for parallel computation on GPU. 
On the other hand, in the step of cluster labeling 
the assignment of label of cluster is done on each site 
piece by piece sequentially; thus the cluster-labeling algorithm 
such as the Hoshen-Kopelman algorithm cannot be directly 
applied to the parallel computation on GPU.

\begin{figure}
\begin{center}
\includegraphics[width=1.0\linewidth]{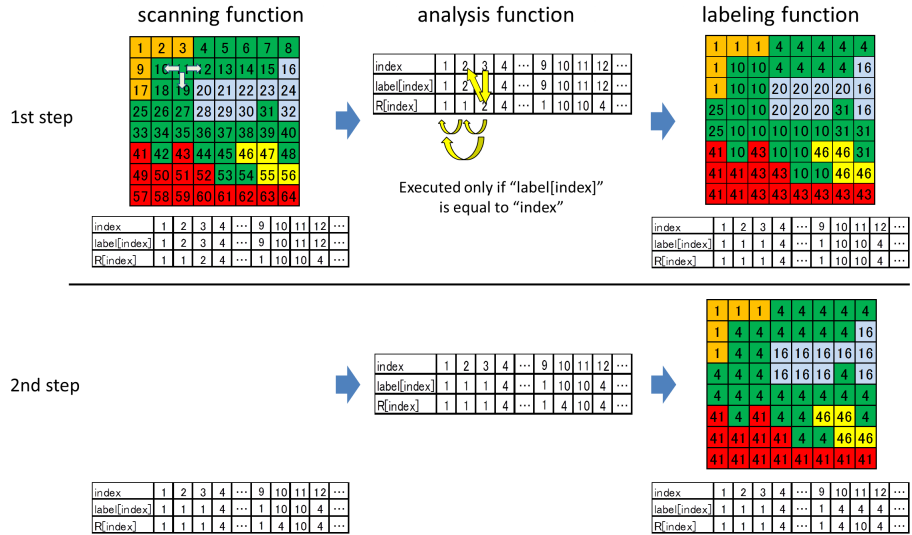}
\caption{\label{fig:fig1} 
Two steps of iterations of the "Label Equivalence" method 
proposed by Hawick {\it et al.} \cite{Hawick_labeling}.  
The connection of sites in the same cluster is represented 
by the same color, and the arrow shows the neighboring sites 
to check for comparison. 
The thread number is denoted by "index".  The variable for saving 
label and the temporal variable are represented by "label" and "R", 
respectively. 
The scanning function compares "label" of each site with 
that of the nearest-neighbor sites. 
If "label" of the nearest-neighbor site is smaller than "label" 
of that site, "R[label]" is updated to the smallest one. 
The equivalence chain of "R" is resolved in the analysis function 
from the starting site to the new site if "label[index]" is equal to "index". 
The labeling function updates "label" 
by label[index] $\leftarrow$ R[label[index]].  Although some clusters 
are not represented by the same "label" at the end of the 1st step 
in this case, all the sites reaches the final label by two steps 
of iteration.  The process of update of "R" in the 2nd step 
is also shown in the figure. 
}
\end{center}
\end{figure}

Recently, Hawick {\it et al.} \cite{Hawick_labeling} 
studied the cluster-labeling algorithm efficient 
for GPU calculation.  Checking four implementations of 
multi-pass labeling method, they proposed the labeling 
method of "Label Equivalence", which is the most efficient 
among four proposals.  
The procedure of their algorithm is explained in figure \ref{fig:fig1}. 
Their algorithm consists of three kernel functions, that is, 
scanning function, analysis function and labeling function, 
and two variables for labeling; one is a variable 
for saving the label, "label" in figure \ref{fig:fig1}, and the other is 
a temporal variable for updated label, "R" in figure \ref{fig:fig1}. 
The scanning function compares the label of each site with 
that of the nearest-neighbor sites when the bond between 
each site and the nearest-neighbor site is active. 
If the label of the nearest-neighbor site is smaller than the label 
of that site, the temporal variable with the label number, 
R[label[index]] in figure \ref{fig:fig1}, is updated to the smallest one. 
For the update of the temporal variable on the scanning function, 
the atomic operation 
\verb+atomicMin()+ is used.  Atomic operations provided by CUDA 
are performed without interference from any other threads. 
The analysis function resolves the equivalence chain of "R" obtained 
in the scanning function; the temporal variable 
\verb+R[index]+ 
is updated from the starting site 
to the new site, which is similar to the method of 
the Hoshen-Kopelman algorithm. 
Each thread checks the temporal variable and the label on each site. 
When the label number, "label", is equal to the thread number, "index", 
each thread tracks back the temporal variable until 
the temporal variable, "R", remains unchanged. 
Since each thread executes this operation concurrently, 
the final value is reached quickly. 
The labeling function updates the label for saving 
by \verb+label[index]+ $\leftarrow$ \verb+R[label[index]]+. 
In the cluster-labeling algorithm due to Hawick {\it et al.}, 
the loop including three functions is iterated up to the point 
when the information of the labeling needs 
no more process of scanning function. 
A small number of iterations 
are needed; 4096$\times$4096 systems with free boundary conditions 
were labeled in 9 or less iterations \cite{Hawick_labeling}.

\begin{figure}
\begin{center}
\includegraphics[width=1.0\linewidth]{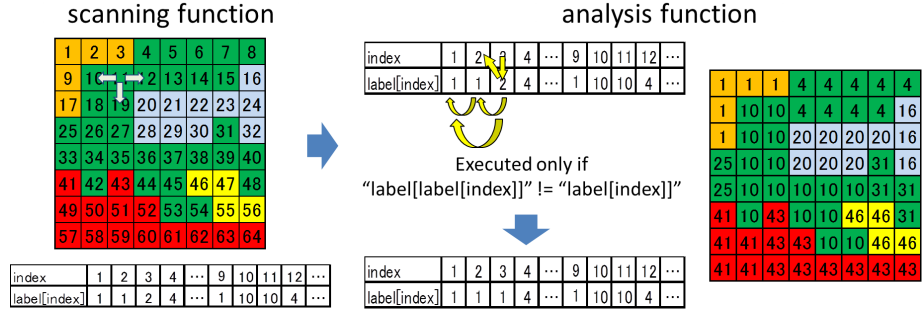}
\caption{
\label{fig:fig2} 
The first step of iterations of the refinement of Label Equivalence 
method proposed by Kalentev {\it et al.} \cite{Kalentev}. 
The meanings of color and arrow are the same as figure \ref{fig:fig1}.
The thread number is denoted by "index". 
The scanning function compares "label" of each site with 
that of the nearest-neighbor sites. 
The equivalence chain is resolved in the analysis function 
from the starting site to the new site if "label[label]" is not equal 
to "label", which results in the update of "label".  
Since the cluster-labeling due to Kalentev is the refined version of 
that due to Hawick, the output of labeling is the same as figure 
\ref{fig:fig1}.  Some clusters are not represented by the same "label" 
at the end of the 1st step in this case, but all the sites reaches 
the final label by two steps of iteration. 
}
\end{center}
\end{figure}

More recently, Kalentev {\it et al.} \cite{Kalentev} reported 
the refinement of the algorithm due to Hawick {\it et al.}. 
The procedure of their algorithm is shown in figure \ref{fig:fig2}.
First, they used only one variable for labeling instead of two 
because there is no need for a temporal reference; 
the implementation was improved in terms of memory consumption. 
It means that the number of kernel functions are reduced 
from three to two because the process of the labeling function 
is no more needed.  
Second, they changed the execution condition on the analysis function 
from "when \verb+label[index]+ is equal to \verb+index+" to 
"when \verb+label[label]+ is not equal to \verb+label+". 
Finally, they eliminated the atomic operation. 
The update of labeling is executed up to the point when the 
labeling needs no more process of the scanning function; 
thus even if collision between threads happens because of 
the absence of the atomic operations, 
it will be resolved during the next iterative step. 
With the refinements due to Kalentev {\it et al.}, the improvement of 
computational speed and the reduction of the memory usage were realized. 

We adapt the two cluster-labeling algorithms, that due to Hawick {\it et al.} 
\cite{Hawick_labeling} and that due to Kalentev {\it et al.} \cite{Kalentev}, 
to the SW multi-cluster algorithm of Monte Carlo simulation. 
In the cluster-labeling algorithms, the label of the cluster is not given 
serially.  To flip the spins in the cluster of the SW algorithm, 
we do not have to know the serial number for the label of the cluster.  
We assign the new spin to any label number even if the cluster of 
that label does not exist.  Because of parallel computation, 
the assignment of new spin to all the possible number of labels 
requires no extra cost. 
To improve the computational speed and save memory, we store 
the information on spin, bond and label in one word; This idea 
was used by Hawick {\it et al.} \cite{Hawick_single_cluster}. 
In the case of treating the system with many spin states, for example, 
we separate the information on spin from the one-word information. 
We finally note that we use a linear congruential random generator  
which was proposed by Preis {\it et al.} \cite{preis09} 
when random numbers are generated. 

\section{Results}
We have tested the performance of our code on NVIDIA GeForce GTX580 
and GTX285. 
For comparison, we run the code on a current CPU, 
Intel(R) Xeon(R) CPU W3680 @ 3.33GHz.  Only one core of the CPU 
is used.  For compiler, we have used gcc 4.1.2 with option -O3. 

We first show the data for the 2D $q$-state Potts models. 
For the cluster-labeling algorithm, we use both the algorithm 
due to Hawick {\it et al.} \cite{Hawick_labeling} and 
that due to Kalentev {\it et al.} \cite{Kalentev}. 
We compare the GPU computational time with the CPU computational time 
at the critical temperature, $T_{c}/J = 1/\ln(1+\sqrt{2}) = 1.1346$ 
for the $q=2$ Potts model (Ising model) and 
$T_{c}/J = 1/\ln(1+\sqrt{3}) = 0.9950$ for the $q=3$ Potts model. 
The average computational times per a spin update 
at the critical temperature for the $q=2$ Potts model 
and the $q=3$ Potts model 
are tabulated in tables \ref{tb:GPU_CPU_time_q=2_Potts} and 
\ref{tb:GPU_CPU_time_q=3_Potts}, respectively. 
There, the time for only a spin update and 
that including the measurement of energy and magnetization are given.
We show the measured time per a spin flip in units of nano sec. 
The linear system sizes $L$ are $L$=256, 512, 1024, 2048 and 4096. 
We can see from tables \ref{tb:GPU_CPU_time_q=2_Potts} and 
\ref{tb:GPU_CPU_time_q=3_Potts} that the computational time of 
our GPU implementation of the SW algorithm is almost constant 
for $L \ge 1024$.  And the computational speed using the algorithm 
of Kalentev {\it et al.} is superior to that of Hawick {\it et al.} 
for all system sizes. 
The performance for $q=2$ with the algorithm of Hawick {\it et al.} is 
2.96 nano sec per a spin flip and 
that with the algorithm of Kalentev {\it et al.} is 
2.51 nano sec per a spin flip with $L=4096$ on GTX580.
The comparison of the performance on GTX580 and that on CPU 
leads to the acceleration of computational speed 
with the algorithm of Kalentev {\it et al.} as 12.4 times for a spin flip 
and 12.6 times for a spin flip with the measurement of energy 
and magnetization for the $q=2$ Potts model with $L=4096$. 
The number of iterations at the critical temperature 
is about 6.6, 7.1, 7.6, 8.1 and 8.6 on average for $L$ = 256, 512, 
1024, 2048 and 4096, respectively; that is, the loop count gradually 
increases with system size.
We here mention the amount of memory used. 
The amount of register is 10 to 13 bytes per thread, and 
the amount of shared memory is 2048 bytes per block for each kernel function. 
These values remain unchanged by system size. 
Using "GPU Occupancy Calculator", we checked that the GPU occupancy is 100\% 
for each kernel function, which indicates that the best performance of GPU 
is attained. 

\begin{table*}[htbp]
\begin{center}
\begin{tabular}{lllllll}
\hline
         &            & $L$=256    & $L$=512    & $L$=1024   & $L$=2048 & $L$=4096    \\
\hline
GTX580                    & update only   & 5.02 nsec & 3.48 nsec & 3.02 nsec & 2.96 nsec & 2.96 nsec\\
\ \ Hawick {\it et al.}   & + measurement & 5.73 nsec & 3.94 nsec & 3.40 nsec & 3.32 nsec & 3.34 nsec\\
GTX580                    & update only   & 4.76 nsec & 3.10 nsec & 2.58 nsec & 2.51 nsec & 2.51 nsec\\
\ \ Kalentev {\it et al.} & + measurement & 5.47 nsec & 3.54 nsec & 2.98 nsec & 2.86 nsec & 2.87 nsec\\
GTX285                    & update only   & 10.0 nsec & 6.96 nsec & 6.14 nsec & 6.03 nsec & 6.04 nsec\\
\ \ Hawick {\it et al.}   & + measurement & 11.2 nsec & 7.63 nsec & 6.70 nsec & 6.55 nsec & 6.60 nsec\\
GTX285                    & update only   & 8.76 nsec & 5.86 nsec & 5.12 nsec & 5.00 nsec & 5.07 nsec\\
\ \ Kalentev {\it et al.} & + measurement & 9.90 nsec & 6.52 nsec & 5.66 nsec & 5.51 nsec & 5.60 nsec\\
Xeon(R) W3680             & update only   & 28.9 nsec & 30.0 nsec & 31.3 nsec & 31.1 nsec & 31.2 nsec\\
                          & + measurement & 33.6 nsec & 34.6 nsec & 36.4 nsec & 36.1 nsec & 36.3 nsec\\
\hline
\end{tabular}
\caption{\label{tb:GPU_CPU_time_q=2_Potts}Average computational time per a
spin flip
at $T_c$ for the $q=2$ Potts model.  The time for only a spin
update and that including the measurement of energy and magnetization are
given.}
\end{center}
\end{table*}

\begin{table*}[htbp]
\begin{center}
\begin{tabular}{lllllll}
\hline
         &            & $L$=256    & $L$=512    & $L$=1024   & $L$=2048 & $L$=4096    \\
\hline                                                     
GTX580                   & update only   & 4.85 nsec& 3.41 nsec& 2.94 nsec & 2.88 nsec  & 2.89 nsec\\
\ \ Hawick {\it et al.}  & + measurement & 5.70 nsec& 3.93 nsec& 3.39 nsec & 3.31 nsec  & 3.31 nsec\\
GTX580                   & update only   & 4.54 nsec& 3.02 nsec& 2.51 nsec & 2.43 nsec  & 2.44 nsec\\
\ \ Kalentev {\it et al.}& + measurement & 5.39 nsec& 3.54 nsec& 2.97 nsec & 2.86 nsec  & 2.85 nsec\\
GTX285                   & update only   & 9.92 nsec& 6.92 nsec& 6.09 nsec & 5.94 nsec  & 5.96 nsec\\
\ \ Hawick {\it et al.}  & + measurement & 11.2 nsec& 7.72 nsec& 6.77 nsec & 6.60 nsec  & 6.61 nsec\\
GTX285                   & update only   & 8.51 nsec& 5.76 nsec& 5.01 nsec & 4.88 nsec  & 4.96 nsec\\
\ \ Kalentev {\it et al.}& + measurement & 9.84 nsec& 6.56 nsec& 5.67 nsec & 5.54 nsec  & 5.60 nsec\\
Xeon(R) W3680            & update only   & 29.2 nsec& 29.2 nsec& 31.5 nsec & 31.4 nsec  & 31.7 nsec\\
                         & + measurement & 35.1 nsec& 34.9 nsec& 37.3 nsec & 37.5 nsec  & 37.6 nsec\\
\hline
\end{tabular}
\caption{\label{tb:GPU_CPU_time_q=3_Potts}Average computational time per a spin flip 
at $T_c$ for the $q=3$ Potts model.  The time for only a spin 
update and that including the measurement of energy and magnetization are given.}
\end{center}
\end{table*}

\begin{figure}
\begin{center}
\includegraphics[width=0.4\linewidth]{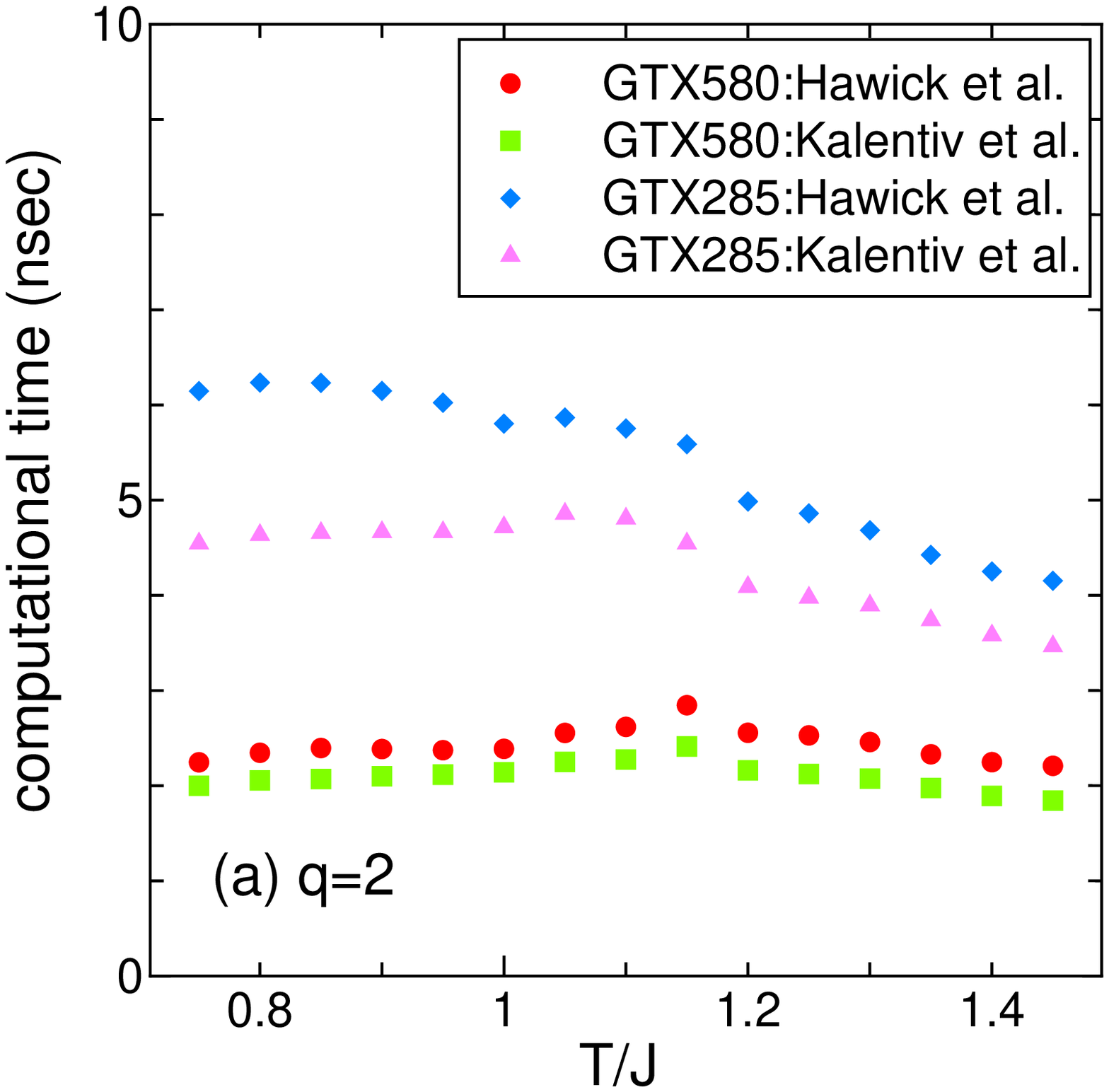}
\hspace{0.02\linewidth}
\includegraphics[width=0.4\linewidth]{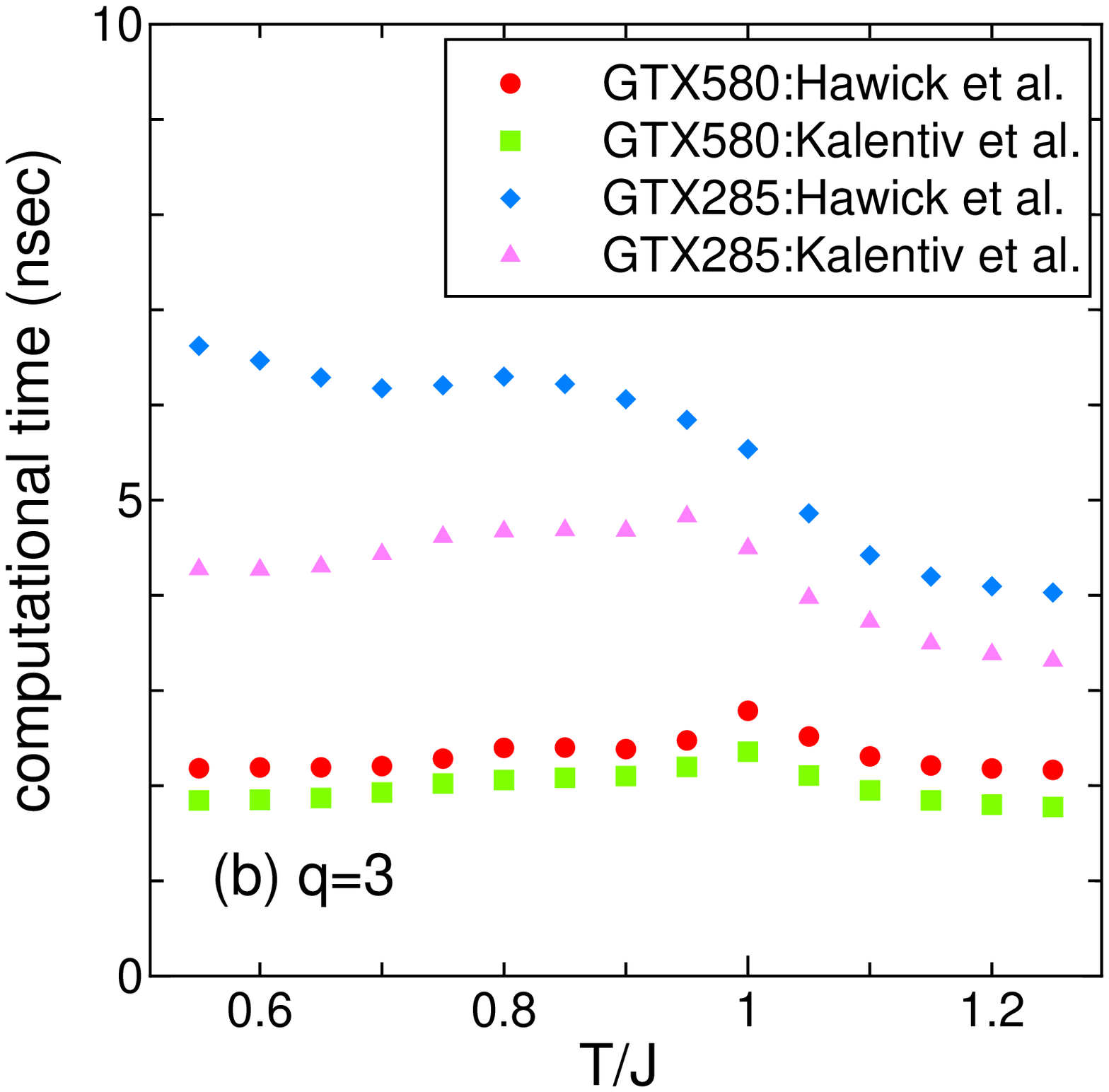}
\caption{\label{fig:fig3}
(a) Temperature dependence of the computational time for GPU computation for the $q=2$ 
Potts model with $L=1024$ and (b) that for the $q=3$ 
Potts model with $L=1024$. }
\end{center}
\end{figure}

Next, we refer to the temperature dependence of our GPU implementation 
of the SW algorithm. 
We plot the temperature dependence of the GPU computational time 
for the $q=2$ Potts model and the $q=3$ Potts model
with $L=1024$ in figures \ref{fig:fig3}(a) and (b), respectively. 
There, we show the average computational time per a spin flip 
with two cluster-labeling algorithms in units of nano sec. 
From figures \ref{fig:fig3}(a) and (b) we can see that 
the computational time is nearly independent 
of temperature. 
Thus, our GPU implementation of the SW algorithm 
is effective for all range of temperatures. 
We observe that 
the computational time becomes a little bit longer near 
the critical temperature, which reflects on the fact 
that the loop count of 
iteration in the cluster labeling increases near the critical 
temperature. 
It may be due to the complex shape of cluster 
near the critical temperature. 

\begin{figure}
\begin{center}
\includegraphics[width=0.4\linewidth]{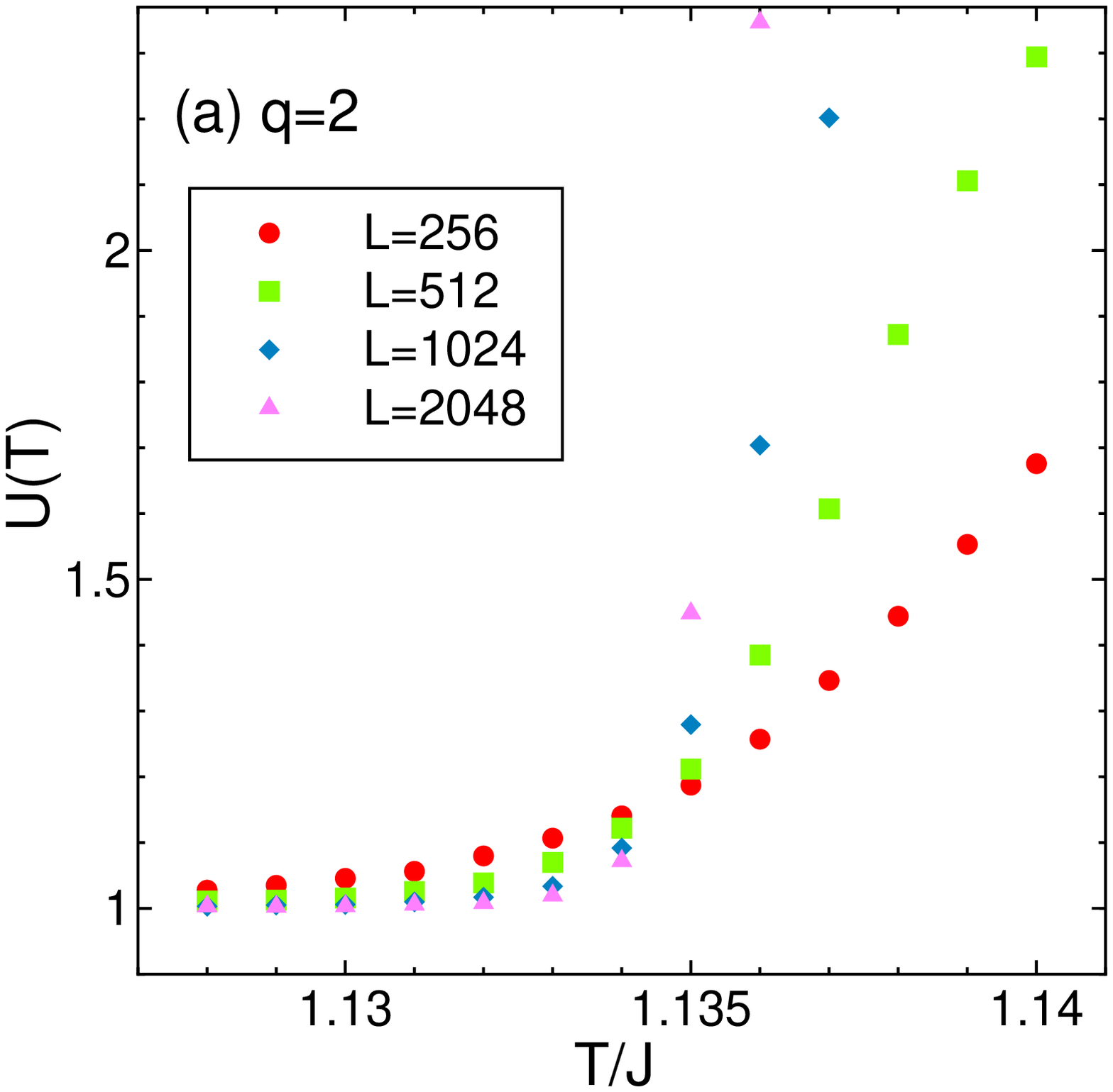}
\hspace{0.02\linewidth}
\includegraphics[width=0.4\linewidth]{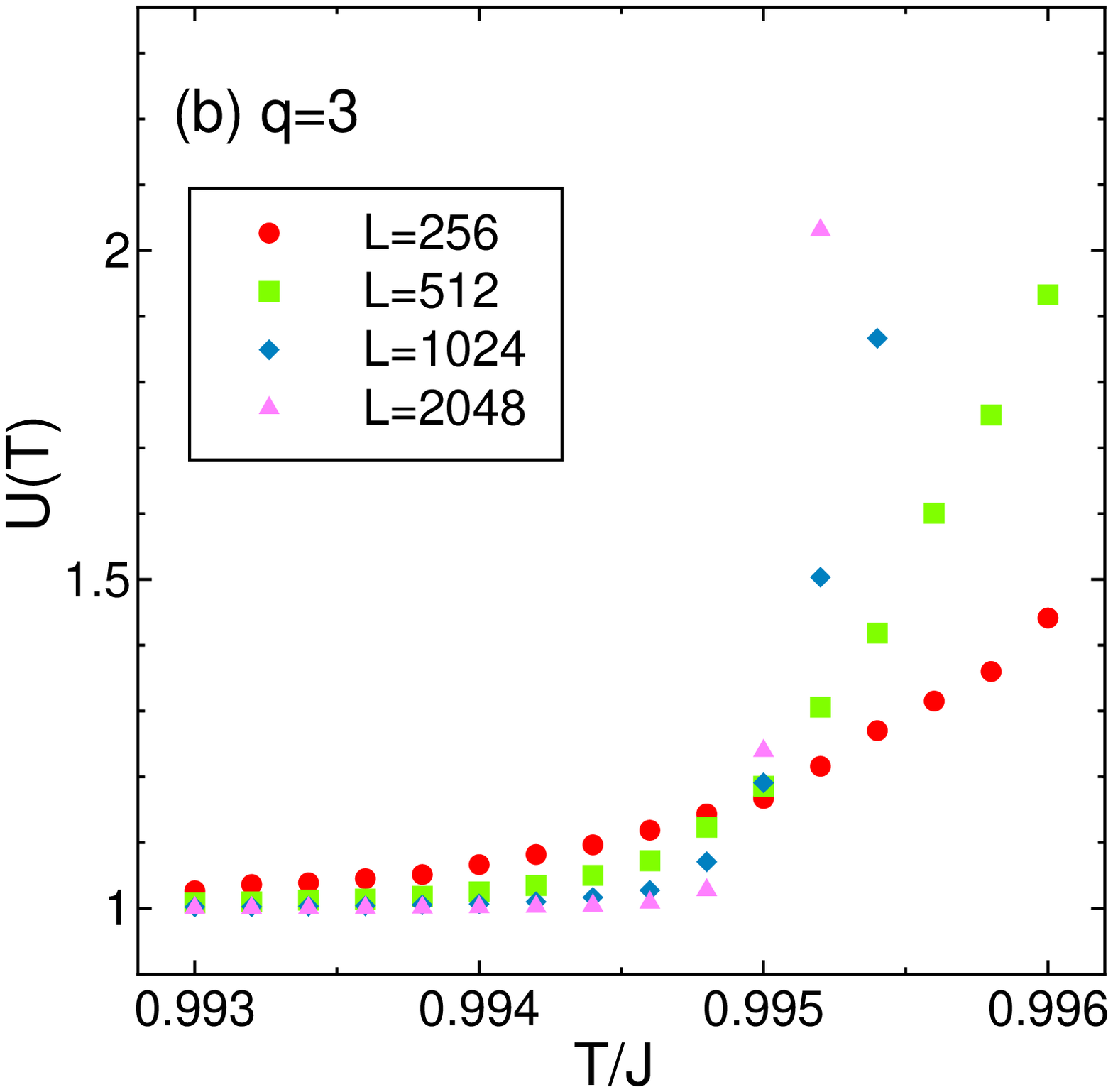}
\caption{\label{fig:fig4}
(a) Moment ratio of the $q=2$ Potts model for $L$=256, 512, 1024 and 2048 and 
(b) that of the $q=3$ Potts model for $L$=256, 512, 1024 and 2048.}
\end{center}

\end{figure}

As an illustration, we plot the moment ratio, 
\begin{equation}
   U(T) = \frac{<M(T)^4>}{<M(T)^2>^2},
\end{equation}
which is essentially the Binder ratio \cite{Binder} 
except for the normalization, 
of the $q=2$ Potts model and the $q=3$ Potts model
in figures \ref{fig:fig4}(a) and (b), respectively. 
The square of the order parameter of the $q$-state Potts model is 
calculated as 
\begin{equation}
   M^2 = \frac{q \ \sum_{k=1}^q n[k]^2-N^2}{q-1},
\end{equation}
where $n[k]$ is the number of spins with the state $k$, and 
$N$ is the total number of spins. 
We here give the data obtained by using the cluster-labeling 
algorithm due to 
Hawick {\it et al.}, for example.  
We discarded the first 10,000 Monte Carlo updates and 
the next 100,000 Monte Carlo updates were used for measurement. 
The crossing of the data with different sizes reproduces 
the known results of the critical temperatures. 

Next, we extend our GPU-based calculation of SW multi-cluster 
algorithm to the system of vector spins.  
We treat the $q$-state clock model, and the Hamiltonian is given by
\begin{equation}
 \mathcal{H} = -J\sum_{<i,j>} \bm{S}_i \cdot \bm{S}_j, 
\end{equation}
where $\bm{S}_i$ is a planar unit vector, 
$(\cos \theta_i, \sin \theta_i)$, at site $i$; 
$\theta_i$ takes the value of $\theta_i = 2\pi p_i/q$ 
with $p_i=1, 2, \cdots, q$.  When $q$ tends to infinity, 
the clock model becomes the classical XY model. 

To make a cluster flip, we use the idea of 
embedded cluster introduced by Wolff \cite{wolff89}. 
We project vector spins to form Ising spin clusters. 
The essential part of the GPU implementation is the same 
as the case of the Potts model. 
We note that the proper use of shared memories is effective 
especially for the calculation of the inner product of vectors. 

As an example, we pick up the $q$-state clock model with $q$=6. 
This model is known to show two Kosterlitz-Thouless 
transitions \cite{KT}, $T_1$ and $T_2$.  
The numerical estimates of $T_1/J$ and $T_2/J$ are 
around 0.7 and 0.9 \cite{tomita2002a}. 
We test the performance of the present implementation 
near the upper critical temperature. 
The average computational time per a spin update 
at $T/J=0.9$ for the $q=6$ clock model 
is tabulated in table \ref{tb:GPU_CPU_time_q=6_up_clock}. 
For the cluster-labeling algorithm, we use both the algorithm 
of Hawick {\it et al.} \cite{Hawick_labeling} and 
that of Kalentev {\it et al.} \cite{Kalentev}.
The computational time for only a spin update and 
that including the measurement of energy and magnetization are given. 
The linear system sizes $L$ are $L$=256, 512, 1024, 2048 and 4096. 
We show the measured time per a spin flip in units of nano sec. 
For the measurement of physical quantities, we also measure 
the correlation function with distances $L/4$ and $L/2$.  
The correlation function is defined as follows: 
\begin{equation}
 G(r,T) = <S_i^xS_{i+r}^x+S_i^yS_{i+r}^y>, 
\end{equation}
and the ratio of the correlation functions with different 
distances is a good estimator for the analysis of 
Kosterlitz-Thouless transition \cite{tomita2002b}.

\begin{table*}[htbp]
\begin{center}
\begin{tabular}{lllllll}
\hline
         &            & $L$=256    & $L$=512    & $L$=1024   & $L$=2048 & $L$=4096    \\
\hline                                                     
GTX580                   & update only    & 4.88 nsec& 3.36 nsec & 2.94 nsec  & 2.85 nsec  & 2.88 nsec\\
\ \ Hawick {\it et al.}  & + measurement  & 6.20 nsec& 4.11 nsec & 3.57 nsec  & 3.44 nsec  & 3.47 nsec\\
GTX580                   & update only    & 4.49 nsec& 2.93 nsec & 2.48 nsec  & 2.38 nsec  & 2.42 nsec\\
\ \ Kalentev {\it et al.}& + measurement  & 5.84 nsec& 3.68 nsec & 3.12 nsec  & 2.98 nsec  & 3.01 nsec\\
GTX285                   & update only    & 10.4 nsec& 7.37 nsec & 6.51 nsec  & 6.21 nsec  & 6.26 nsec\\
\ \ Hawick {\it et al.}  & + measurement  & 12.7 nsec& 8.76 nsec & 7.68 nsec  & 7.32 nsec  & 7.36 nsec\\
GTX285                   & update only    & 8.65 nsec& 5.86 nsec & 5.15 nsec  & 4.97 nsec  & 5.09 nsec\\
\ \ Kalentev {\it et al.}& + measurement  & 10.9 nsec& 7.25 nsec & 6.32 nsec  & 6.08 nsec  & 6.19 nsec\\
Xeon(R) W3680            & update only    & 83.4 nsec& 83.2 nsec & 84.5 nsec  & 86.4 nsec  & 86.3 nsec\\
                         & + measurement  & 99.4 nsec& 108.6 nsec& 114.5 nsec & 124.7 nsec & 128.8 nsec\\
\hline
\end{tabular}
\caption{\label{tb:GPU_CPU_time_q=6_up_clock}Average computational time per a spin flip 
at $T/J=0.9$ for the q=6 clock model.  The time for only a spin 
update and that including the measurement 
of energy, magnetization and correlation function with distances $L/4$ and $L/2$ are given.}
\end{center}
\end{table*}

Although the calculation of the clock model on CPU takes 
much more time than that of the Potts model, 
the computational time for the GPU-based calculation 
of the clock model is almost the same as that 
of the Potts model.  The proper use of shared memories 
may contribute to the good performance for the clock model. 
The performance of the $q=6$ clock model with 
the cluster-labeling algorithm of Hawick {\it et al.} 
is 2.88 nano sec per a spin flip and 
that with the algorithm of Kalentev {\it et al.} is 
2.42 nano sec per a spin flip with $L=4096$ on GTX580. 
The acceleration of computational speed over the 
calculation on CPU with the algorithm 
of Kalentev {\it et al.} is 35.6 times for a spin flip 
and 42.7 times for a spin flip including the measurement 
of energy, magnetization 
and correlation function with distances $L/4$ and $L/2$ 
for the $q=6$ clock model with $L=4096$. 

\begin{figure}
\begin{center}
\includegraphics[width=0.5\linewidth]{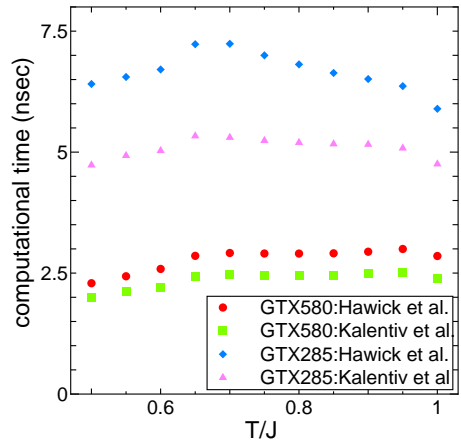}
\caption{\label{fig:fig5}
Temperature dependence of the computational time for GPU computation for the $q=6$ 
clock model with $L=1024$. }
\end{center}
\end{figure}

The temperature dependence of our GPU-based calculation 
of the SW algorithm for the $q=6$ clock model is plotted 
in figure \ref{fig:fig5}. The linear system size is $L=1024$. 
We show the average computational time per a spin flip 
in units of nano sec.  From figure \ref{fig:fig5} we can see that 
the computational time weakly depends on the temperature. 
Thus, we can say that our GPU implementation of the SW 
multi-cluster algorithm is also effective for the clock model 
in all range of temperatures. 

\begin{figure}
\begin{center}
\includegraphics[width=0.8\linewidth]{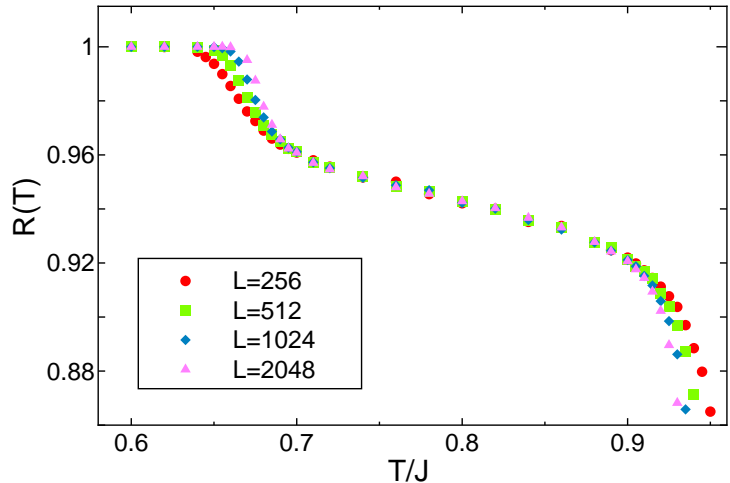}
\caption{\label{fig:fig6}
Temperature dependence of the ratio of the correlation function for the $q=6$ clock model for $L$=256, 512, 1024 and 2048. }
\end{center}
\end{figure}

As an illustration, we plot the ratio of the correlation function 
\begin{equation}
 R(T) = \frac{G(L/2,T)}{G(L/4,T)}
\end{equation}
of the $q=6$ clock model in figure \ref{fig:fig6}. 
We discarded the first 10,000 Monte Carlo updates and 
the next 400,000 Monte Carlo updates were used for measurement. 
From figure \ref{fig:fig6}, we see that the curves of different 
sizes overlap in the intermediate Kosterlitz-Thouless phase 
($T_1<T<T_2$), and spray out for the low-temperature ordered 
and high-temperature disordered phases.  
The graph reproduces the result shown in Fig. 2 of 
Ref.~\cite{tomita2002b} for small sizes.
Recently, the estimate 
of two transition temperatures of $q=6$ clock model 
is an issue of controversy \cite{hwang,baek}.  
The detailed analysis of the clock models 
using the finite-size scaling analysis 
will be given elsewhere. 

\section{Summary and discussion}
We have formulated a GPU parallel computing of the SW multi-cluster 
algorithm by using the two connected component 
labeling algorithms, the algorithm by Hawick {\it et al.} 
\cite{Hawick_labeling} and that by Kalentev {\it et al.} 
\cite{Kalentev}, for the assignment of clusters. 
Starting with the $q$-state Potts model, we also extended 
our implementation to systems of vector spins 
using the idea of embedded cluster by Wolff \cite{wolff89}. 
We have tested the $q$-state Potts models with $q$=2 and 3 
and the $q$-state clock model with $q=6$ by use of our implementation 
of the SW algorithm. 
As a result, the GPU computational time by using the 
cluster-labeling algorithm by Kalentev {\it et al.} is 
2.51 nano sec per a spin update for the $q=2$ Potts model 
and 2.42 nano sec per a spin update for the $q=6$ clock model 
on GTX580 with the linear size $L=4096$ at the critical temperature. 
The performance of the algorithm by Kalentev {\it et al.} 
is superior to that of Hawick {\it et al.} for all models and sizes. 
It confirms the effectiveness of refinement by Kalentev {\it et al.}; 
that is, the elimination of atomic operation and 
the reduction of the number of kernel functions. 
We obtained that the computational time of our implementation 
is almost constant for the linear size $L \ge 1024$, 
and there is little temperature dependence 
for our SW multi-cluster algorithm. 

Now we compare the performance of our implementation of the 
SW multi-cluster algorithm with that of Weigel \cite{weigel11}. 
He uses the combination of self-labeling algorithm and 
label relaxation algorithm or hierarchical sewing algorithm. 
Comparing with the breadth-first search and the tree-based 
union-and-find approach, the self-labeling algorithm is used 
in partitioning a set of elements into disjoint subsets.  
To consolidate cluster labels, 
the label relaxation algorithm and the hierarchical sewing algorithm 
are used. 
The GPU computational time of his algorithm was reported as 
2.70 nano sec per a spin update for the $q=2$ Potts model at 
the critical temperature with the linear size $L=8192$ 
on GTX580.  
The performance of the algorithm of Weigel strongly depends 
on system size and temperature, and this speed of 2.70 nano sec 
per a spin update is reached only for $L=8192$.  
The performance becomes much worse for $L<8192$ and 
at temperatures below the critical temperature. 
On the other hand, 
the GPU computational time of our algorithm 
is 2.51 nano sec for the same model with $L=4096$ 
on the same GPU, GTX580, 
and our implementation of the SW algorithm 
has little dependence on system size and temperature.

We have shown the data up to $L=4096$ in this paper 
because we use one-dimensional index in launching a CUDA kernel. 
Since the amount of memory on GTX580 is 1.5 Gbyte, 
we can treat the system up to $L=8192$ by using the two-dimensional index. 
The algorithm employed here implements the labeling 
over the whole lattice instead of partitioning. 
Because of the flexibility of our implementation, 
it will be interesting to apply the present formulation 
to multi-GPU calculations.  
There are advantages in both our implementation and that by Weigel. 
The application of GPU to cluster algorithms has just started. 
This problem deserves further attention.

\section*{Acknowledgment}
This work was supported by a Grant-in-Aid for Scientific Research from
the Japan Society for the Promotion of Science.

\end{document}